\input harvmac.tex

% Something to deal with sub-sub-sections

\def\unlockat{\catcode`\@=11}

\def\lockat{\catcode`\@=12}

\unlockat

% Something to deal with sub-sub-sections

\def\newsec#1{\global\advance\secno by1\message{(\the\secno. #1)}
\global\subsecno=0\global\subsubsecno=0
\global\deno=0\global\prono=0\global\teno=0\eqnres@t\noindent {\bf\the\secno. #1}
\writetoca{{\secsym} {#1}}\par\nobreak\medskip\nobreak}
\global\newcount\subsecno \global\subsecno=0
%%%%%%%%%%%%%%%%%%%%%%%%%%%%%%%%%%%%%%%%%%%%%%%%%%%%%%%
\def\subsec#1{\global\advance\subsecno
by1\message{(\secsym\the\subsecno. #1)}
\ifnum\lastpenalty>9000\else\bigbreak\fi\global\subsubsecno=0
\global\deno=0\global\prono=0\global\teno=0%\eqnres@t\noindent
\noindent{\it\secsym\the\subsecno. #1} \writetoca{\string\quad {\secsym\the\subsecno.} {#1}}
\par\nobreak\medskip\nobreak}
\global\newcount\subsubsecno \global\subsubsecno=0
%%%%%%%%%%%%%%%%%%%%%%%%%%%%%%%%%%%%%%%%%%%%%%%%%
\def\subsubsec#1{\global\advance\subsubsecno by1
\message{(\secsym\the\subsecno.\the\subsubsecno. #1)} \ifnum\lastpenalty>9000\else\bigbreak\fi
\noindent\quad{\secsym\the\subsecno.\the\subsubsecno.}{#1}
\writetoca{\string\qquad{\secsym\the\subsecno.\the\subsubsecno.}{#1}}
\par\nobreak\medskip\nobreak}
%%% Definition

\global\newcount\deno \global\deno=0
\def\de#1{\global\advance\deno by1
\message{(\bf Definition\quad\secsym\the\subsecno.\the\deno #1)}
\ifnum\lastpenalty>9000\else\bigbreak\fi \noindent{\bf
Definition\quad\secsym\the\subsecno.\the\deno}{#1}
\writetoca{\string\qquad{\secsym\the\subsecno.\the\deno}{#1}}}
%%% Proposition

\global\newcount\prono \global\prono=0
\def\pro#1{\global\advance\prono by1
\message{(\bf Proposition\quad\secsym\the\subsecno.\the\prono #1)}
\ifnum\lastpenalty>9000\else\bigbreak\fi \noindent{\bf
Proposition\quad\secsym\the\subsecno.\the\prono}{#1}
\writetoca{\string\qquad{\secsym\the\subsecno.\the\prono}{#1}}}
%%% Theorem

\global\newcount\teno \global\prono=0
\def\te#1{\global\advance\teno by1
\message{(\bf Theorem\quad\secsym\the\subsecno.\the\teno #1)}
\ifnum\lastpenalty>9000\else\bigbreak\fi \noindent{\bf
Theorem\quad\secsym\the\subsecno.\the\teno}{#1}
\writetoca{\string\qquad{\secsym\the\subsecno.\the\teno}{#1}}}
%%%%%%%%%%%%
\def\subsubseclab#1{\DefWarn#1\xdef
#1{\noexpand\hyperref{}{subsubsection}%
{\secsym\the\subsecno.\the\subsubsecno}%
{\secsym\the\subsecno.\the\subsubsecno}}%
\writedef{#1\leftbracket#1}\wrlabeL{#1=#1}}% Macros for boxes

\lockat

%%%%%%%%%%%%%%%%%%%%%  Rublenye bukvy   %%%%%%%%%%%%%%%%%%%%%%%%

\def\IB{\relax\hbox{$\inbar\kern-.3em{\rm B}$}}
\def\IC{\relax\hbox{$\inbar\kern-.3em{\rm C}$}}
\def\ID{\relax\hbox{$\inbar\kern-.3em{\rm D}$}}
\def\IE{\relax\hbox{$\inbar\kern-.3em{\rm E}$}}
\def\IF{\relax\hbox{$\inbar\kern-.3em{\rm F}$}}
\def\IG{\relax\hbox{$\inbar\kern-.3em{\rm G}$}}
\def\IGa{\relax\hbox{${\rm I}\kern-.18em\Gamma$}}
\def\IH{\relax{\rm I\kern-.18em H}}
\def\IK{\relax{\rm I\kern-.18em K}}
\def\IL{\relax{\rm I\kern-.18em L}}
\def\IP{\relax{\rm I\kern-.18em P}}
\def\IR{\relax{\rm I\kern-.18em R}}
\def\IZ{\relax\ifmmode\mathchoice
{\hbox{\cmss Z\kern-.4em Z}}{\hbox{\cmss Z\kern-.4em Z}} {\lower.9pt\hbox{\cmsss Z\kern-.4em
Z}} {\lower1.2pt\hbox{\cmsss Z\kern-.4em Z}}\else{\cmss Z\kern-.4em Z}\fi}

\def\II{\relax{\rm I\kern-.18em I}}

\def\frac#1#2{{#1\over#2}}
%%%%%%%%%%%%%%%%%%%%% Calligraphic letters  %%%%%%%%%%%%%%%%%%%%%

%%%%%%%%%%%%%%%%%%%%%%%%%% Derivatives  %%%%%%%%%%%%%%%%%%%%%%%%

%%%%%%%%%%%%%%%%%%%% letters with bar %%%%%%%%%%%%%%%%%%%%%%%%%%

%%%%%%%%%%%%%%%%%%%%%%%%%%% Math symbols %%%%%%%%%%%%%%%%%%%%%%%

\def\inbar{\,\vrule height1.5ex width.4pt depth0pt}
\font\cmss=cmss10 \font\cmsss=cmss10 at 7pt

%%%%%%%%%%%%%%%%%%%%%%%%%%%%%%%%%%%%%%%%%%%%%%%%%%%%%%%%%%%%%%%%%

\font\manual=manfnt \def\dbend{\lower3.5pt\hbox{\manual\char127}}

% Macros for boxes

\def\boxit#1{\vbox{\hrule\hbox{\vrule\kern8pt
\vbox{\hbox{\kern8pt}\hbox{\vbox{#1}}\hbox{\kern8pt}} \kern8pt\vrule}\hrule}}
\def\mathboxit#1{\vbox{\hrule\hbox{\vrule\kern8pt\vbox{\kern8pt
\hbox{$\displaystyle #1$}\kern8pt}\kern8pt\vrule}\hrule}}

%%%%%%%%%%%%%%%%%%%%%%%%%%%%%%%%%%%%%%%%%%%%%%%%%%%%%%%%%%%%%%%%%%%%
%%%%%%%%%%%%%%%%%%%%%%%%%%%%%%%%%%%%%%%%%%%%%%%%%%%%%%%%%%%%%%%%%%%%
%%%%%%%%%%%%%%%%%%%%%%%%%%%%%%%%%%%%%%%%%%%%%%%%%%%%%%%%%%%%%%%%%%%%
\Title{ \vbox{\baselineskip12pt \hbox{hep-th/0011009} \hbox{YCTP-SS1-2000 } \hbox{ITEP-AG-2000}
\hbox{} }} {\vbox{
 %\centerline{}
\centerline{Stringy Higgs Mechanism }
%\bigskip \centerline{Open String Field Theory}
\bigskip
\centerline{and}
\bigskip
 \centerline{the Fate
of Open Strings}}}
\medskip
\centerline{\bf A. A. Gerasimov $^{1}$, and S. L. Shatashvili $^{2}$\footnote{*}{On leave of
absence from St. Petersburg Branch of Steklov Mathematical Institute,
 Fontanka,
St. Petersburg, Russia.}}

\vskip 0.5cm \centerline{\it $^{1}$ Institute for Theoretical and Experimental Physics, Moscow,
117259, Russia} \centerline{\it $^{2}$ Department of Physics, Yale University, New Haven, CT
06520-8120 }
%\centerline{}

\vskip 1cm We propose a refinement of the physical picture describing different vacua in
bosonic string theory. The vacua with closed strings and open strings are connected by the
string field theory version of the Higgs mechanism, generalizing the Higgs mechanism of an
abelian gauge field interacting with a complex scalar. In accordance with Sen's conjecture, the
condensation of the tachyon is an essential part of the story. We consider this phenomenon from
the point of view of both a world-sheet sigma-model and the target-space theory. In the
Appendix the relevant remarks regarding the choice of the coordinates in the background
independent open string field theory are given.

\medskip
\noindent

\Date{November 1, 2000}

\newsec{Introduction}

Since the conjectures regarding the vacuum structure of open string theory were made by Sen
\ref\sen{A. Sen, Tachyon condensation on the brane antibrane system, hep-th/9805170, JHEP 9808
(1998) 012.}, \ref\sentwo{A. Sen, Universality of the Tachyon Potential, hep-th/9911116, JHEP
9912 (1999) 027.}, considerable interest has been devoted to its verification as well as study
of the consequences. The studies of \ref\gs{A. A. Gerasimov and S. L. Shatashvili, On exact
tachyon potential in open string field theory, hep-th/0009103, JHEP 0010:034,2000.},
\ref\greg{D. Kutasov, M. Marino and G. Moore, Some exact results on tachyon condensation in
string field theory, hep-th/0009148.}, \ref\sennew{D. Ghoshal and A. Sen, Normalisation of the
background independent open string field theory, hep-th/0009191.} demonstrated that the
approach based on background independent open string field theory of \ref\wbi{E. Witten, On
Background Independent Open String Field Theory, hep-th/9208027, Phys. Rev. D46 (1992)
5467-5473; Some Computations in Background Independent Open-String Field Theory,
hep-th/9210065, Phys. Rev. D47 (1993) 3405-3410.}, \ref\shatashone{S. Shatashvili, Comment on
the Background Independent Open String Theory, hep-th/9303143, Phys. Lett. B311 (1993) 83-86;
On the Problems with Background Independence in String Theory, Preprint IASSNS-HEP-93/66,
hep-th/9311177, Algebra and Anal., v. 6 (1994) 215-226.} can be very effective in addressing
the question of tachyon condensation since one can find the exact expression for the string
field theory action order by order in a derivative expansion of space-time fields. In this
standard way of treating the low energy lagrangian, the exact answer already in the two
derivative approximation provides important information and allows one to demonstrate the
validity of Sen's conjectures.\foot{It is very interesting to note that in \ref\minzw{J.
Minhahan and B. Zwiebach, Field theory models for tachyon and gauge field string dynamics,
hep-th/0008231, JHEP 0009:029,2000; Effective tachyon dynamics in superstring theory,
hep-th/0009246.}, motivated by rather different ideas, the same tachyon lagrangian, both in the
bosonic and superstring case, was proposed as a toy model that mimics the expected properties
of tachyon condensation.} Further generalizations were considered in \ref\co{L. Cornalba,
Tachyon condensation in large magnetic fields with background independent open string field
theory, hep-th/0010021.}, \ref\ok{K. Okuyama, Noncommutative tachyon from background
independent open string field theory, hep-th/0010028.} for the case of a constant $B$-field and
in \ref\gregtwo{D. Kutasov, M. Marino and G. Moore, Remarks on tachyon condensation in
superstring field theory, hep-th/0010108.} for the case of superstrings.

According to the Sen's conjectures for the open bosonic string theory, the process of tachyon
condensation leads to a new vacuum state which doesn't contain any open string states. It is
naturally identified with the perturbative vacuum of the closed bosonic string theory. This
gives the convincing support to the idea that {\it the perturbation theories of open and closed
strings are expansions in some background independent universal theory around the different
vacua}.

We would like to propose a further explanation of the connection
 between open and closed strings, guided by the symmetries
 of string theory. It turns out that in the
 process of learning how the
 electro-magnetic field disappears as a result of tachyon
 condensation in the approach of background independent open string field
 theory, we come to an important understanding
 regarding the properties of string theory in general:
 the connection between the two vacua is a close relative
of the standard  Higgs description of two types of
 vacua  in the theory of an abelian gauge field coupled to
a complex scalar field. The perturbative vacuum of the closed strings plays the role of the
invariant vacuum, and the perturbative vacuum of the open strings is the vacuum with the
spontaneously broken symmetry.

In the vacuum with the spontaneously broken symmetry the open
string tachyon is similar to the absolute value of the complex
scalar field in the field theory example, and the role of the
phase is played by the other fields in the open string spectrum.
In this analogy the gauge transformations of the open string
fields are similar to the identification of the angular variable
$\theta \sim \theta +2\pi$. This is the manifestation of the
particular choice of the coordinates and not of the topology of
the configuration space. As the consequence, for the other
coordinates (e.g.
 natural variables in the closed string theory)
there is no trace of these open
 string gauge symmetries. This serves as the explanation of
 the absence of the open string gauge fields after tachyon
 condensation. We note that the tachyon potential in open string
field theory is very different from the standard field theory example
$V(\phi)=(|\phi|^2-\eta^2)^2$.
 In particular the vacuum with the spontaneously broken symmetry
is unstable. Thus the perturbative expansion around the symmetric vacua is in terms of closed
string degrees of freedom. The other degrees of freedom
 are suppressed in the symmetric vacuum by the infinite effective mass.
 From the point of view of the closed string theory world-sheets,
 these hidden degrees of freedom correspond to non-smooth
 world-sheet deformations.

The main suggestion we make in this paper is closely related to the search for closed strings
inside the field theory of open strings; this has been a challenge for many years (see e.g.
\ref\str{A. Strominger, Closed Strings in Open String Field Theory, Phys. Rev. Lett., v. 58, n.
7, (1987) 629.}) and more evidence has occurred recently \ref\sam{S. Shatashvili, {\it Closed
strings as solitons in background independent open string field theory}, unpublished, talk at
IHES, Paris, July 1997.} in the view of developments related to D-branes and Matrix Theory
\ref\bfss{T. Banks, W. Fischler, S. Shenker and L. Susskind, M-theory as Matrix Model: A
Conjecture, hep-th/9610043, Phys.  Rev.D55  :5112-5128, 1997.}.\foot{It has been proposed in
\sam\ (based on several evidences extracted from Matrix Stings on D1 brane in IR limit) that in
background independent open string field theory (D9 or D25) there should exist a solitonic
solution corresponding to fundamental, Nambu-Goto, closed string and thus open string field
theory can serve as the definition of full, self-consistent theory of interacting open and
closed strings.} Note also that a related qualitative picture of the disappearance of the open
string gauge fields and the emergence of the closed strings was presented in \ref\senfield {A.
Sen, Some Issues in Non-commutative Tachyon Condensation, hep-th/0009038.} and \ref\sennew{A.
Sen, Fundamental Strings in Open String Theory at the Tachyonic Vacuum, hep-th/0010240.}.

In section 2 we consider the truncated version of the open/closed string theory (the
approximation we use is rather close to the considerations in \ref\witt2{E. Witten, AdS/CFT
Correspondence And Topological Field Theory, hep-th/9812012, JHEP 9812:012,1998 }). We
demonstrate that the process of the tachyon condensation closely follows the Higgs mechanism in
the theory of a $U(1)$-gauge field interacting with a complex scalar (of course with an
appropriate map between fields of abelian Higgs model and truncated string field theory). Based
on this analogy we give a qualitative picture of the new vacuum. The considerations of this
section suggests that one should think about the Higgs phenomena taking place in the space of
loops instead of space-time (when all stringy modes are included). If the gauge bundles over
the
 space-time play a prominent role in the  usual quantum theories
 of fields, the proper analogs in the theory of strings  are gerbes with
 2-connections.  Thus we
 suggest that the basic mechanism behind the open/closed string
 transformation is, in the nutshell, the Higgs phenomenon for gerbes.

 The fact that the $B$ field absorbs the gauge field in the Higgs mechanism, familiar
 from
 supergravity theories, was know from early days of string theory \ref\csh{E. Cremmer, J. Scherk, Spontaneous
dynamical
 breaking of gauge symmetry in dual models,
    Nucl.Phys.B72:117-124,1974.},
 and in the context of Born-Infeld action
 \ref\chiara{C.G. Callan, C. Lovelace, C.R. Nappi, S.A. Yost, String loop
 corrections to
 beta function,
  Nucl. Phys .B288: 525, 1987. }
  (for some early studies of tachyon condensation see
  \ref\marty{K. Bardakci,  Dual Models and Spontaneous Symmetry Breaking, Nucl. Phys.
B68 (1974) 331; K. Bardakci and M. B. Halpern, Explicit Spontaneous Breakdown in a Dual
Model, Phys. Rev. D10 (1974) 4230; K. Bardakci and M. B. Halpern,
Explicit Spontaneous Breakdown in a Dual
Model II: N Point Functions, Nucl. Phys. B96 (1975) 285;
K. Bardakci, Spontaneous Symmetry Breakdown in the Standard Dual String
Model, Nucl. Phys. B133 (1978) 297.}).
The appearance of
the gauge group
$SU(N)$ instead of $U(N)$
  in the AdS/CFT correspondence
   \ref\wit3{
   E. Witten, Anti de Sitter Space And Holography,
   hep-th/9803002, Adv. Theor. Math. Phys. 2: 253-291, 1998.} possibly
 has the same origin.

 It is well known that the separation of closed string and open string degrees of freedom
 is rather ambiguous. The inclusion of the singular interactions in the open string theory gives
 the closed string states in the loop expansion. On the other hand, one could use
 the smooth open string vertices
 at the cost of explicit introduction of closed string fields.   We claim: after tachyon is properly included in
the picture
 and when it condenses, the coordinates with explicit closed strings become more appropriate
  and  closed strings  appear as dynamical variables. We conjecture
 that all open string fields (except tachyon) are ``angle'' variables and
 correspond to gauge parameters for corresponding closed string fields.

In section 3 we begin to analyze this interpretation  directly in the sigma model approach. The
transformation of the open strings (2d surfaces with boundaries) to the closed strings (2d
surfaces without boundaries) is a "geometric" one and this leads us to believe that the sigma
model approach could provide the basic insight into the question of the disappearance of open
strings.
 In the new vacuum, the condensate of the
  tachyon zero mode becomes  infinite, and due to the general
  prefactor $e^{-RT_0}$ for each open string loop,  forbids
a non-zero boundary on the world-sheet.\foot{Related question was asked by S. Shenker and we
thank him for sharing it with us.} In this section we discuss the connection between these two
vacua in terms of the 2d quantum theory on the world-sheet and find the picture of spontaneous
symmetry breaking similar to the one discussed in section 2.

In order to gain more information in section 4 we use the open/closed string field theory due
to Zwiebach \ref\zwieb{B. Zwiebach Oriented Open-Closed String Theory Revisited,
hep-th/9705241, Annals Phys. 267: 193-248, 1998. } to verify the off-shell connection between
open string fields and closed string gauge parameters.

In the Appendix we discuss two important questions regarding the
role of the choice of the proper coordinates in the background
independent open string field theory action.

To conclude this introduction we would like to mention that the question: {\it what is exactly
the space of boundary field theories in 2d?} seems to be of the fundamental importance in the
open/closed string relation.\foot{This question have been raised by E. Witten over the years
since \wbi, \shatashone.} In this notes we give just first steps towards the possible answer.

\newsec{Stringy Higgs mechanism: target-space approach}

In order to study the role of higher spin fields in the process of tachyon condensation we
start with the truncated field content
 in the open/closed string theory.
 Thus we consider the metric $G$,
 two-form gauge field $B$, open string $U(1)$ gauge field $A$ and open
 string tachyon $T$.

  Let us remark that in general the inclusion
 of the full spectrum of open and closed strings in the action
 leads to the over-counting of the degrees of freedom and their
contributions
 to amplitudes. But here we are
 looking at the truncated open string theory and this
 consideration is legitimate.

In background independent open string field theory we start from the world-sheet theory defined
on the disk with the operator $e^{-\int d\theta (T+AdX + ...)}$ inserted on the boundary (we
assume that ghosts decouple). This operator is invariant under open string gauge
transformations $A \rightarrow A+d\Lambda$ and global shifts $A \rightarrow A + const$. The
space-time action for these fields has the form (see the Appendix for the choice of the
parameterization):
 \eqn\oneaction{S(G,B,A,T)=S_{closed}(G,B)+\int d^{26}X \sqrt{G}
 (e^{-T}(1+T)+e^{-T}||dT||^2+{1\over 4}e^{-T}||B-dA||^2+\cdots )}
One shall note the obvious gauge invariance of the action:
 \eqn\gaugeB{B\rightarrow B+da}
 \eqn\gaugeA{A\rightarrow A+a}
 where $a$ is 1-form gauge parameter. Global symmetry corresponds to constant $a$.

We can consider two situations: closed string modes are fixed
(non-dynamical) backgrounds and  closed string modes are
dynamical. As it follows from the tachyon potential, in both cases
we have two types of the vacua in the  theory. At the open string
 perturbative vacuum, $T=0, A=0$, we have the tachyon of finite mass  and
therefore
 this vacuum is unstable. In the theory of only open strings we also have the massless gauge field $A$.
 When we make the closed string
modes dynamical (which is not necessary at the moment) - in this vacuum the gauge invariance
\gaugeB , \gaugeA\ allows to put
 $A$ to be zero and we are left with the tachyon and
mass term for the field $B$.

It appears that the open string tachyon potential has another vacuum for the infinite value of
the tachyon field $T=\infty$ (in addition there are many soliton solutions corresponding to the
lower D-branes but we will not discuss them here). In the new variables:
\eqn\changenew{\Sigma=e^{-{1\over 2}T}} it is at $\Sigma=0$ (unstable vacuum is at $\Sigma=1$)
and:

\eqn\oneaction{S(G,B,A,\Sigma)=S_{closed}(G,B)+\int d^{26}X \sqrt{G}
 (\Sigma^2(1-2\log \Sigma)+4||d\Sigma||^2+{1\over
4}\Sigma^2||B-dA||^2+\cdots )}

Around the critical point $\Sigma=0$ from corresponding potential
in \oneaction\  we conclude that the square of mass for
$\Sigma$-field (``tachyon'') is positive and infinite. According
to Sen this should be the vacuum corresponding to the theory of
the closed strings. Note that the kinetic term for the gauge field
$A$ multiplies zero, $\Sigma^2$, at this point and thus the gauge
field is not well defined at this vacuum.

Last expression \oneaction\ immediately suggests the analogy with the quantum field theory
textbook Lagrangian: \eqn\twoaction{S(\Phi,{\cal A})=\int dX
 ({1\over g^2}F({\cal A})^2+ |d\Phi-i{\cal A}\Phi|^2+\lambda
(|\Phi|^2-|\Phi_0|^2)^2)} with gauge transformations:
 \eqn\abeliantr{{\cal A} \rightarrow {\cal A} + d\chi }
 \eqn\abeliantrone{ \Phi \rightarrow e^{i\chi}\Phi}
that are analogous to \gaugeB, \gaugeA. Here we have the abelian gauge field $\cal A$
interacting with the complex scalar field $\Phi$ and as an example the forth order polynomial
potential.

In order to make the connection to string theory lagrangian \oneaction\ we allow ourselves to
briefly review the properties of \twoaction\ in the variables similar to \oneaction:

\eqn\radialpar{\Phi(X)=e^{i\phi(X)}H(X)} The action \twoaction\ is: \eqn\twoactiontwo{S(H,
\phi,{\cal A})=\int dX
 ({1\over g^2}F({\cal A})^2+ H^2|d\phi-{\cal A}|^2+|dH|^2+\lambda
(H^2-H_0^2)^2)} In the new coordinates, $\phi$ is identified under the shift transformation:
 \eqn\shiftr{\phi \rightarrow \phi +2\pi}
The appearance of the  field identification \shiftr\ is not the manifestation of the
non-trivial topology of the configuration space, but it is the artifact of the choice of the
special coordinate system. For instance in terms of two scalar fields $X,Y$ ($\Phi=X+iY$) there
is no condition like \shiftr.

In these variables the gauge transformations are exactly like for string theory case \gaugeB,
\gaugeA:
 \eqn\abeliantr{{\cal A} \rightarrow {\cal A} + d\chi }
 \eqn\abeliantrone{ \phi \rightarrow \phi +\chi}
The stable vacuum $\Phi=\Phi_0$ is not invariant under phase-shift
- it is stable and non-symmetric. When gauge field ${\cal A}$ is
background field, we have massive scalar $H$ and massless scalar -
phase $\phi$. When ${\cal A}$ is dynamical - we can set the
angular variable to zero by gauge transformations and we get
massive $H$ and massive gauge field $A$.

There is also unstable but symmetric vacuum - $\Phi=0$. In radial
variables for background gauge field ${\cal A}$ we have tachyonic
field $H$ and the phase field $\phi$ is ill-defined since its
kinetic term multiplies zero expectation value of $H$-field.  This
is the manifestation of the fact that these variables are not well
defined at this point and one should use another (non-singular)
parameterization (e.g in terms of $X$ and $Y$). At the same time,
when gauge field becomes dynamical we can perfectly live with
angular variables; the angular field $\phi$ is a gauge parameter
for ${\cal A}$ - we get massive tachyonic field $H$ plus massless
gauge field ${\cal A}$. This is how field theory abelian Higgs
mechanism looks in angular coordinates.

The analogy between \twoactiontwo\ and \oneaction\ is quite obvious. We can ``map'' the
variables as:

\eqn\mapone{\Sigma \rightarrow H} \eqn\maptwo{A \rightarrow \phi} \eqn\mapthree{B \rightarrow
{\cal A}} The description of open string theory in terms of the fields $\Sigma, A$ in
\oneaction\ is similar to the description of abelian field theory in terms of matter fields $H,
\phi$. Closed string mode $B$ is related to gauge field ${\cal A}$. The tachyon (more exactly
the field $\Sigma$) plays the role of the "radial" component of the complex scalar field and
the gauge field is the analog of the "angular"
 variable. The gauge field $A$ shifts under the gauge transformation
\gaugeA\ similar to the $\phi$ in the previous considerations. It is interesting that the usual
gauge transformation of $A$ is the full analog of the shift symmetry \shiftr.

The analogy described above taught us the following: we can study
the truncated string theory system around the vacuum $(T=0
,\Sigma=1), A=0$ (unstable and non-symmetric) and
 the corresponding situation in abelian Higgs model is
stable/non-symmetric vacuum. Otherwise, if we study the string theory around the new vacuum at
$\Sigma=0$ which is invariant with respect to closed string  symmetry \gaugeB\
(``stable''/symmetric vacuum) - it is similar to the symmetric point in the configurations
space of the abelian gauge field theory example (unstable/symmetric). But from the field theory
considerations we know that the description of the fluctuations around symmetric vacuum
configuration in terms of the "angular" type variables and fixed background field ${\cal A}$ is
totally inappropriate and one should use the different parameterization of the fields (for
example cartesian). Now, we shall note that latter problem in abelian Higgs model is removed by
choosing the correct coordinates, and by dynamical gauge field ${\cal A}$. If we go back to
string theory example it seems that angular variables are forced on us from the beginning in
truncated open/close string system and in case of fixed closed string background there is no
possibility to introduce the analog of local cartesian coordinates $X, Y$ since the analog of
absolute value of $\Phi=H$ (which is $\Sigma$) and phase factor $\phi$ (which is $A$) carry
different space-time spin. At the same time we can introduce non-local string field theory
wave-function \eqn\wilson{\Psi(X(\sigma))=e^{-\int(T(X(\sigma)+A(X)dX(\sigma)+\cdots )d\sigma}}
which may be considered as the formal analog of the complex scalar field $\Phi$ and the 2-form
$B$ field gives the natural connection on the space of these functionals (this means that we
now need to include all fields and all derivatives in space-time lagrangian which becomes the
lagrangian in loop space). Thus we conclude that in the truncated string model it is forced to
use non-local string field variable and assume that closed string modes are dynamical. The
latter is an important conclusion since what has been claimed is that there is a new branch in
open string field theory where the only dynamical degrees of freedom are closed strings. Two
branches are connected by new, stable, closed string vacuum $\Sigma=0$.

All this support the idea that at the new vacuum of the string theory  there is no open string
states in the spectrum and we have the theory of the closed strings instead.

It is  interesting to note that the string field theory wave function \wilson\  is closely
related  to cubic ${\bf CS}$ open string field theory coordinates (see \gs). Let us remember
that in \gs\ the following relation between tachyon modes in sigma-model and cubic {\bf CS}
string field theory coordinates was proposed: consider world-sheet path integral for the disk
topology and divide the disk into two equal parts with first half carrying the fixed boundary
condition $X(z, \bar z)|_{\partial D}=X_*(\sigma)$ and on second half the operator \wilson\
inserted. For tachyon zero modes this leads to the relation:
 \eqn\comsigma{e^{-{1\over 2}T_0}=1+T_0^{CS}}
We see that the cubic ${\bf CS}$ string field theory expansion is built around the vacuum
corresponding to $T_0^{CS}=0$. One can use the conformal transformation in order to map the
disk to ``$1/3$ of pizza'' - with $120^o$ angle segment and glue three such wave-functions in
order to get familiar cubic term in {\bf CS} action which now becomes the disk partition
function with functional \wilson\  inserted on the boundary (second term in the background
independent action $S=-\beta^i \partial_i Z + Z$); as far as the $\beta$-function term in the
action - it is obviously obtained from the kinetic term $\Psi Q \Psi $ of {\bf CS} cubic action
(for the appropriate regularisation in sigma model).

\newsec{World-sheet considerations}

Now we would like to understand the above picture in terms of
world-sheet sigma model. Consider the perturbation series
expansion in the string theory with open strings. We should sum up
the contributions from the arbitrary genus surfaces with arbitrary
number of the holes. The contribution of each surface is given by
the 2d functional integral over the fields with Neumann boundary
conditions (``partition function'') which has the expansion:
 \eqn\actiontotal{Z_{Total}=\sum _{h,n} {1\over
n!}g^{2h-2+n}Z_{\Sigma_{g,n}}}

We will mainly be interested in the contributions of the open string loops and thus restrict
ourselves by the genus zero surfaces.

\eqn\actionone {Z_*=\sum _{n} {1\over n!}g^{n-2}Z_{\Sigma_{0,n}}}

This expansion may be interpreted as the closed string partition
function in the "shifted" background. Being not very precise one
could say that there is an operator in 2D theory such that its
insertion in the correlator simulates the appearance of the hole
on the world-sheet. If we denote this operator $V_H$ then the
partition function \actionone\ may be formally written as an
``effective action'' for the closed string in some new background:
 \eqn\actionloop{Z_*=\sum_{n} {1 \over n!}<{V_H}^n>=<e^{V_H}>}

The addition of this operator to the
 world-sheet action
deforms the vacuum of the closed string theory to the new vacuum where the world-sheets with
holes are possible.

This operator may be described  explicitly. Consider Hilbert space of states of the quantum
fields defined on the circle of the radius $R$  in the first quantized closed string theory.
The vacuum state is define in terms of  the annihilation operators in the standard way:
 \eqn\ahihilat{\alpha^i_{n}|vac>=0}
Geometrically it means that this vacuum state is induced by the
functional integral over the disk. This state is invariant under
various symmetries. For instance it is invariant under the gauge
transformation of the 2-form field $B\rightarrow B+da$ which will
be important in the following consideration. The perturbative
expansion around this vacuum is given in terms of the standard
closed string modes. Corresponding operators are given by the
polynomials over creation operators (up to the momentum factor
$e^{ip_iX^i}$).

In open string diagrams one chooses  Neumann boundary conditions - the normal derivative of the
fields on the boundary should vanish. This is very different from the previous case and doesn't
correspond to the vacuum state induced by the integral over the disk. Thus in this case we have
the  real hole on the string world sheet. The vanishing normal derivative is equivalent to the
condition:
 \eqn\conditone{\partial_{\sigma} X^i_L(\sigma)=\partial_{\sigma}
 X^i_R(\sigma)}
 Here $X^i_{L,R}$ are restrictions on the boundary of the chiral and antichiral
 components of the scalar fields. The corresponding condition on
 the open string vacuum state may be conveniently written in terms
 of the canonical momentum variables $P^i_{n}={1\over 2}\int d\sigma (\partial_{\sigma} X^i_L(\sigma)
 -\partial_{\sigma} X^i_R(\sigma))$ as:
 \eqn\conditvac{P^i_n|open>=0}

 The transformation between two states is a standard
 Bogolubov transformation with some operator:
 \eqn\boper{|open>=U_H|vac>}
  \eqn\boperone{U_H=e^{{1\over 2}A(\alpha ,\alpha)}}
where  $A(\alpha ,\alpha)$ is a bi-linear form on the creation and annihilation operators. Its
explicit expression is not important at the moment.

The corresponding wave function:
 \eqn\wavefunction{\Psi(X(\sigma))=<X(\sigma)|open>}
may be considered as the classical solution of the closed string theory corresponding to the
open string vacuum. To get the corresponding sigma model vertex operator we should recall that
the additional moduli of conformal structure appears when we consider the surface with the cut
disk instead of the puncture. This additional parameter may be identified with the radius of
the disk and the corresponding vertex is  naturally one- differential that should be integrated
over this parameter to get the 'hole cutting operator" that we are looking for:
 \eqn\den{V_H=\int dR U_H(R)}

This description uses the explicit parameterization of the moduli space of conformal
structures. One could propose the invariant definition of this operator which does not use the
explicit parameterization. Consider the contour $L$ on the 2d surface and let $V_L$ be the
operator which force the fields in the functional integral to have zero normal derivative on
the contour $L$. The the coordinate independent analog of \den\ would be given by the integral
over the contours:

 \eqn\diffinvar{V_H^{inv}=\int dL V_h(L)}

 Perturbations around this vacuum are quite different from
the closed string states and may be naturally described in terms of open string states. Note
also that this new vacuum is not invariant with respect the symmetries of the closed string
vacuum. In particular the gauge symmetry of the $B$ field does not leave it invariant. This
leads to the conclusion that in this new vacuum we have the spontaneous breaking of the
symmetry and thus some kind of Higgs type effect.

The coordinates in the vicinity of this new vacuum, as usual, have a flavor of the radial
coordinates. There are degrees of freedom connected with the symmetries of the theory. The most
obvious example is the open string abelian gauge field which is in a sense
 one of the parameters of the closed string gauge (BRST)
 transformations. These fields are similar to the angular
 variables. In particular they are defined up to some
 identification. In the case of the standard angular variable it
 is the identification:
 \eqn\identone{\theta\sim \theta +2\pi}
while in the case of the gauge fields it is a gauge transformation:
 \eqn\gaugeident{A\sim A+d\phi}

Obviously this gauge symmetry is an artifact of the parameterization and shows up only around
the non-trivial vacuum.

The role of the radial coordinates plays the open string tachyon which is invariant with
respect to the closed string gauge transformations. In the open string  vacuum the expectation
value of every "radial" variable is non-zero and  it is tempting to conclude that the operator
\diffinvar\  is just the open string constant tachyon operator.

The presented picture rises the following question. In the previous section we have argued that
the open string states become infinitely massive in the closed string vacuum. But here we move
in the opposite direction and looking for the open strings in terms of perturbative closed
string theory. We believe  that the answer is in the subtleties of the short-distance
description of the world-sheet. The open string modes which are infinitely massive in the
closed string vacuum are responsible for the non-smooth deformations of the world-sheets and
thus are formally absent (have infinite mass) in the perturbative first quantized closed string
states. The insertion of the hole cutting  operator just change the space of states drastically
and the new degrees of freedom are brought about. Note in connection with this the whole
question of open/closed string transformation is known to be deeply related with short-distance
behaviour of the world-sheet QFT (see e.g. \ref\strom{A. Strominger, Phys.Lettt. 58 (1987)629;
Nucl.Phys. B294 (1987)93}).

Let us remark that the description of the open string vacuum in terms of the summation over the
surfaces is not quite appropriate. We are looking at the theory in the "shifted" vacuum (open
strings) using basically the description in terms of world-sheets natural for another vacuum
(closed strings).

What is the right framework for the expansion around open string vacuum is not quite clear. The
only hint we have is the AdS/CFT correspondence where we are looking at the regime when $V_H$
operator dominates in the action
 \ref\verlind{J. Khoury, H. Verlinde, On Open/Closed String
Duality, hep-th/0001056.}

 One could consider these two vacua from another point of view.
Consider the open string theory in the background of the constant
open string tachyon mode $T=T_0$. The 2d theory is not conformal
due to the term on the boundary. This boundary term would give the
factor for each boundary:

\eqn\aone{Z\sim e^{-RT_0}}
 where $R$ is the length of the boundary.

According to the sigma-model approach one should fix the conformal factor of the metric and
integrate over the moduli of conformal structure. At the solutions of the equation of motions
the answer does not depend on the choice of the conformal factor due to conformal invariance
and away from conformal point it can be compensated by the  field-redefinition. Thus we could
fix the conformal factor as we want. Let us take some boundary to be of unit length, then we
have the overall damping factor of the string amplitudes with the holes. Thus we may conclude
that at the new vacuum $T_0=\infty$ and there should not be any boundary at all.

In terms of above mentioned "hole cutting" operator on could say that at the new string vacuum
$T=\infty$ the coefficient in front of this operator in the 2D action is zero. Thus there are
no holes and no open strings in this vacuum.

\newsec{String Field Theory approach.}

The connection of the open string fields with the parameters of
the gauge transformations of the closed strings was important part
of the arguments presented above. In this section we test this
connection in the framework of the open/closed string field theory
constructed by Zwiebach \zwieb. We will follow closely the
notations of \zwieb.

The basic structure used in this construction is Batalin-Vilkovisky (BV) algebra. This
structure make possible the correct quantization of the theory. Let $\Psi$ and $\Phi$ be closed
string and open string wave functions. To define the structure of BV algebra one should define
the odd brackets coming from the odd symplectic form and $\Delta$ operator acting on this
functionals. It was demonstrated in \zwieb\ that one could use the product structure on the
space of open and closed string functionals:
\eqn\aaone{\{A,B\}_{o/c}=\{A,B\}_{open}+\{A,B\}_{closed}}

The full action: \eqn\aatwo{S(\Phi,\Psi)=S(\Phi)_{o}+S(\Psi)_c+S(\Phi,\Psi)_{int}} should
satisfy the full quantum  BV-equation:

\eqn\bv{\hbar \Delta S+{1\over 2}\{S,S\}=0}

Now consider the classical properties of the BV geometry. The
classical part of this equation is:
 \eqn\bvclass{\{S,S\}=0}
Given the odd symplectic structure on the space one has the
algebra of symplectic transformations leaving this symplectic
structure invariant. Locally these transformations are generated
by the hamiltonian functions $U$:
 \eqn\hamilt{\delta A=\{A,U\}}
In particular these transformations leave  invariant the condition
\bvclass\ and thus transforms the action functional in the field
theory into another one suitable for quantization. In general it
is not the gauge transformation of the theory because the action
functional is not necessary invariant. Gauge transformations are
given by the hamiltonian functions of the special form:
 \eqn\gaugeham{U=\{f,S\}}
 where $f$ is an arbitrary function. It is easy to verify that the
 invariance of the action functional is the direct consequence
 of the equation \bvclass .

The general structure of the action functional for the open/closed
string theory may be described as follows. The action functional
for the open string has the form:
 \eqn\actionopen{S_o(\Phi)={1\over 2}<\Phi, Q
 \Phi>+<\Phi^3>+<\Phi^4>+\cdots}
where the first term is the free open string action. It is the
solution of the open string BV equation (closed strings dropped
out):

 \eqn\bvopenone{\{S_o,S_o\}_o=0}
It defines the structure of $A_{\infty}$ algebra on the open string functionals.

Similarly, the closed string action has the form: \eqn\actionclosed{S_o(\Psi)={1\over 2}<\Psi,
Q
 \Psi>+<\Psi^3>+<\Psi^4>+\cdots}
and is the solution of the closed string BV equation:
 \eqn\bvopenone{\{S_c,S_c\}_c=0}
(The $L_{\infty}$ structure on the closed strings.)
 The important point for our discussion is the appearance of the
 following second  term in the interaction part of the action:
  \eqn\interact{S(\Psi,\Phi)_{int}=<\Psi> +<\Psi |\Phi> +<\Psi |
  \Phi^2>+\cdots}

It was shown in \zwieb\ that this coupling between the close and
open strings  is essential to get the integral over the full
moduli space of conformal structures of the surfaces with
boundaries in the perturbative expansion of the string field
theory.

Consider now the a gauge transformation with the following
hamiltonian function:
 \eqn\maingauge{U=\{<\eta |\Psi>,S\}}
where the $\eta$ is the closed string ghost one state. Taking into account the expressions for
the parts of the action \actionopen, \actionclosed , \interact\ we have
 \eqn\explgauge{U=-<Q\eta |\Psi>+<\eta|\Phi>+\cdots}

Here dots are instead of various non-linear terms. Now it is
obvious that the gauge symmetry of the full open/closed string
action has the form:
 \eqn\aaaone{\delta \Phi=\{\Phi,U\}=(\eta)_o+\cdots }
  \eqn\aaatwo{\delta \Psi=\{\Psi,U\}=-Q\eta+\cdots }

We introduce the notation $(\eta)_o$ here to stress that it is the
non-trivial projection of the closed string sector to the open
string sector.

>From the transformations \aaaone\ and \aaatwo\ we may conclude
that there is the gauge symmetry in the theory of open and closed
strings which is the gauge transformation in the closed string
sector and the shift in the open string sector. The whole
machinery of BV formalism guarantees that this first order
transformations could be correctly completed up to the full
non-linear symmetry of the theory.

At the end we would like to note that in the framework of \zwieb\ the arguments about the
suppressing the boundaries by the factor $e^{-RT_0}$ mentioned previously becomes well defined.
Just because all string vertexes used in \zwieb\ have  stubs at the limit $T_0\rightarrow
\infty$ closed string part hopefully makes a leading contribution.

\newsec{On the possible generalizations}

One could look at the tachyon in the closed string theory from the same perspective. There
should be  a new "conformal point"
 at $T\rightarrow \infty$. At the new vacuum the 2d surfaces are
 summed with the coefficient:
 \eqn\gone{Z\sim e^{-T_0 A}}
 where $A$ is the area of the surface. Probably the same arguments
 lead to the conclusion that 2d surfaces should shrink to the
 point and we leave with closed 3d surface (if any).

 One could test these ideas in the case of analog of
 M-theory and "little" string theories. In the latter case
 we have 2-branes with the boundaries on the other branes.
 And the condensation of the tachyon may lead to annihilation
 of these branes. \foot{In a sense this is an analog of \wbi\ for
 closed strings -  boundary of (2+1)d membrane; the possibility that the
closed
 string field theory could be constructed in analogous to \wbi\ fashion
 via the 3d surface with boundary was suggested long ago \ref\ver{E.
Verlinde, E. Witten,
 private communications, 1993.}.}

\vskip 1cm

{\bf Acknowledgments:} We would like to thank T. Appelquist, J. Bagger, J. Maldacena, N.
Nekrasov, H. Ooguri, S. Sachdev, J. Schwarz,  S. Shenker, L. Takhtajan, E. Verlinde and E.
Witten for important discussions. The research of A.G. is partially supported by RFBR grant
98-01-00328 and the research of S. Sh. is supported by OJI award from DOE.

\newsec{\bf Apendix}

Here we would like to discuss some of the question of regularisation dependence (choice of
coordinates in the space of fields) for background independent open string field theory
lagrangian on the example of tachyon. The importance of the right choice of the coordinates to
have the metric on the fields of the canonical form was already stressed in the main part of
the text. Expansion of partition function in the derivatives of tachyon field and gauge field
was performed long ago in the original paper on sigma model approach  in \ref\fts{E. Fradkin
and A. Tseytlin, Quantum String Theory Effective Action,  Nucl.  Phys.,  B261, 1-27 (1985).}.
Here we will more closely follow the line of the reasoning in \gs. We start with the
description of the coordinates, used in \gs\ for the derivation of the tachyon action up to two
derivative terms and then consider the generalization to the case of the abelian gauge fields.
In this approximation the general expressions for $\beta$-function an the partition function
may be written as follows:
 \eqn\betagen{\beta^T (X)=a_0(T)+a_1(T)\partial T+a_2(T)\partial^2
 T +a_3(T)(\partial T)^2+\cdots }
 \eqn\partgen{Z=\int d^{26}Xe^{-T(X)}(1+b(T)(\partial T)^2+\cdots
 )}
To make considerations more simple we begin with natural coordinates in the sigma model
approach and find corresponding $\beta$ and $Z$. In these coordinates the boundary action has
the form $\int d\sigma T(X)$.

 It is rather obvious that zero mode of the tachyon enters in the
partition function as the overall prefactor $e^{-T_0}$. Therefore we may infer that
$b(T)=const$, $a_0(T)=T, a_1=const, a_2=const, a_3=const$ and $T_0$ enters the $\beta$-function
as an additive term:
 \eqn\betatwo{\beta_T(T)=T_0+(T_0-independent\, terms)}

 The partition function and $\beta$-function for the quadratic
profile $T=a+{ u \over 4} X^2$ were calculated in \wbi .
 Using these explicit calculations we
conclude that: $a_0(T)=T, b=a_1=a_3=0, a_2=2$.

Substituting the $\beta$-function: \eqn\betawit{\beta_T(T)=T+2\Delta T} and partition function:
 \eqn\partgen{Z=\int d^{26}Xe^{-T(X)}}
  in the
basic equation: \eqn\baseq{S=-\beta^i\partial_i Z+Z} we have the following action:
 \eqn\actionone{S = \int dX^{26} e^{-T} [2 (\partial T)^2 + (T + 1)]}
with the corresponding equations of motion:
 \eqn\EOMone{e^{-T}(T+4\Delta T-2(\partial T)^2)=0}
The equations of motion are related to the $\beta$-functions by the  metric $G_{ij}$ on the
space of fields.

 \eqn\connection{\partial_i S = G_{ij} \beta^j}
Comparing  \EOMone\  with the  \betawit\ we find:
 \eqn\compareone{G(\delta_1 T,\delta_2 T)=\int dX
 e^{-T}(\delta_1T\delta_2 T-2(d\delta_1 T)(d\delta_2 T))}

This metric is rather complex and not obviously is an expansion of some invertible metric on
the space of fields.
Now we make a change of coordinates leading to more simple form of the
metric (note that these new coordinates were used in \gs\ in order to write down the tachyon
action exactly following the above line of reasoning).

We make this field redefinition in two steps.

  First we consider the linear $T$-term. Note that $T$-linear term
in $\beta$ function is invariant with respect to the scaling of the tachyon field
($\beta^i\partial_i$ is a vector field) while the linear part of the equation of motions (or
quadratic part of the action) is not.

Consider the new coordinates on the space of the tachyon configurations:
 \eqn\newcoord{T\rightarrow T-\partial^2 T}
We have the following expressions:

\eqn\newbeta{\beta_T(X)=T+2\Delta T}
 \eqn\newpart{Z(T)=\int e^{-T}(1+(\partial T)^2+\cdots)}
 \eqn\newaction{S = \int dX e^{-T} (T+1)((\partial T)^2 + 1)}
Obviously, in terms of \wbi\ this is equivalent to different
 regularisation of the Green function at coincident points.

Now the linear terms in the equation of motion are proportional to the linear terms in the
$\beta$-function with the simple coefficient $e^{-T}$.  Note that it is obviously invertible
for finite $T$.

Finally consider the  new coordinate $T_*$:
 \eqn\newone{T_*= T-(\partial T)^2}
In terms of the new coordinate we have ($b=0$):
 \eqn\zone{Z(T_*)=\int dX e^{-T_*}(1+\cdots)}
The corresponding $\beta$-function may be obtained from the condition of the invariance:
 \eqn\betaone{\beta(T(X)){\delta \over \delta T(X)}=
 \beta_*(T_*(X)){\delta \over
\delta T_*(X)}}  and has the following form:

\eqn\betatwo{\beta_*(T_*)=T_*+2\Delta T_*-(dT_*)^2} From this we obtain the final action (in
the new coordinates $T_*$): \eqn\newtwo{S = \int dX e^{-T_*} [(\partial T_*)^2 + (T_* + 1)]}
with the equations of motion: \eqn\great{ {\partial \over {\partial T_*}} S(T_*) =   e^{-T}
(T_*+2\Delta T_*-(dT_*)^2) = e^{-T_*} \beta(T_*) =0} This is the form of the  action given in
\gs.

Thus we have demonstrated  that the action \newtwo\ is connected with the action \actionone\
 by the field redefinition:
 \eqn\redefinition{T\rightarrow T-\partial^2 T +(\partial
 T)^2+\cdots}
 and that in the variables \newtwo\ the metric which relates equations of
motion
 and world-sheet $\beta$-function has the simple form of multiplication by
$e^{-T_*}$
 as opposed to the one for \actionone.

Now let us include into the consideration the abelian gauge field. Using the same approximation
we consider the part of the lagrangian which depends polynomially on the gauge field
stress-tensor (up to second order) and does not depend on its derivatives. This approximation
is similar to taking into account the first two terms of the expansion of  Born-Infeld
lagrangian.

At this approximation we have the following expressions for the partition function and
$\beta$-functions generalizing \betagen\ and
\partgen

\eqn\betagenem{\beta^T (X)=T+2(T)\Delta T+c_1(T)F^2 \cdots }
 \eqn\betagenemm{\beta^A(X)_{\nu}=c_2\partial^{\mu}TF_{\mu \nu}+\cdots}
 \eqn\partgenem{Z=\int d^{26}Xe^{-T(X)}(1+c_3(T)F^2+\cdots  )}
Analogously to the case of the pure tachyon we could deduce that $c_i$'s are independent on
$T$.

Using the definition of the action functional \baseq\  one  has in the necessary approximation:

 \eqn\actionnewem{S(T,A)=\int dX^{26} e^{-T}((T+1)+2(\partial
T)^2+(c_1+c_3)F^2+c_3T F^2+\cdots )} with the corresponding equations of motion  for the
tachyon being:
 \eqn\EOMnew{{\delta S \over \delta T}=-T-2\Delta T-c_1F^2-c_3TF^2+\cdots
} Consider new tachyon field:
 \eqn\newtachyon{ T^*=T+c_3F^2}
 \eqn\newem{A^*=A}
Taking into account the property of the covariance for the beta function:
 \eqn\covaem{\beta^T {\partial \over \partial T}+
\beta^A {\partial \over \partial A}=\beta^{T^*} {\partial \over
\partial {T^*}}+\beta^{A^*} {\partial \over \partial A^*}}
we find that the new beta function is:
 \eqn\newbetaem{\beta^{T^*}(X)=-T^*-2\Delta T^*+(c_1+c_3)F^2}
Thus, in the new coordinates the action and the equations of motion are:

 \eqn\actionnewemone{S(T^*,A^*)=\int dX^{26} e^{-T^*}((T^*+1)+2(\partial
T)^2+(c_1+c_3)F^2+\cdots )}

 \eqn\EOMnewone{{\delta S \over \delta T^*}=-T^*-2\Delta
T^*-(c_1+c_3)F^2+\cdots }

Combining   \redefinition,
\newtachyon\ and \newem\  we have the lagrangian:

 \eqn\actionnewemfinal {S(T,A)=\int dX^{26} e^{-T}((T+1)+(\partial
T)^2+(c_1+c_3)F^2+\cdots )} The coefficients $c_1=0$ and $c_3={1\over 4}$ obtained in the
appropriate regularisation scheme could be read off from the formula (3.22) in \ref\Keke{K. Li
and E.Witten, Role of Short Distance Behavior in Off-shell Open-String Field Theory,
hep-th/9303067, Phys.Rev. D48 (1993) 853-860.} (see also \co\ and \ok). Covariantizing the
action with respect to diffeomorphisms  and $B_{\mu \nu}$ field gauge transformations we end up
with the following result:
 \eqn\actionnewemfinal{S(T,A)=\int
dX^{26}\sqrt{G} e^{-T}((T+1)+||d T||^2+{1\over 4} ||B-F||^2+\cdots )} One shall note that the
choice of coordinates as in lagrangian \actionnewem\ for $c=0, c_3={1 \over 4}$ would lead to
gauge field dependence through $\int V(T) F^2$ which is natural to expect since it is second
order term in expansion of Born-Infeld action replacing the metric in $\int V(T) \sqrt G$. Our
principle of choice of coordinates leads to \actionnewemfinal\ instead.

\listrefs

\bye